# Multiwavelength Photoluminescence as a Non-destructive Probe of Depth-Dependent UV Damage in Perovskite Semiconductors

Zixu Huang,[1] Julisa Juarez,[1] Akash Dasgupta,[1] David S. Ginger[1,*]

1. Department of Chemistry, University of Washington, Seattle, WA, 98195 USA

* Corresponding author: dginger@uw.edu

**Abstract**

Buried interfaces play a critical role in the operational stability of perovskite optoelectronic devices, yet remain challenging to probe non-destructively. Here, we demonstrate a multiwavelength photoluminescence-based approach exploiting the wavelength-dependent absorption of perovskites, enabling fast, and non-destructive probing of depth-sensitive degradation behavior of perovskite devices under operational conditions. Using ultraviolet (UV) light-soaking with different indium tin oxide (ITO) modifiers as a test case, we employ wide-field photoluminescence (PL) imaging, time-resolved photoluminescence (trPL) and ideality factor measurements with different wavelength excitation to characterize degradation behavior of both perovskite thin films and photovoltaic devices under UV soaking. Using only non-destructive characterization, we show that UV-induced degradation in p-i-n perovskite solar cells occurs preferentially at the HTL interface, adversely affecting device stability, consistent with ex-situ destructive tests. Our results also show that modifying the buried interface with 3-mercaptopropyltrimethoxysilane reduces UV-induced damage at the buried interface, thus improving the UV stability of both perovskite films and devices. This work establishes a fast, non-destructive strategy to screen depth-dependent degradation in perovskite devices, offering insights for interface design and stability optimization.

**TOC graphic**

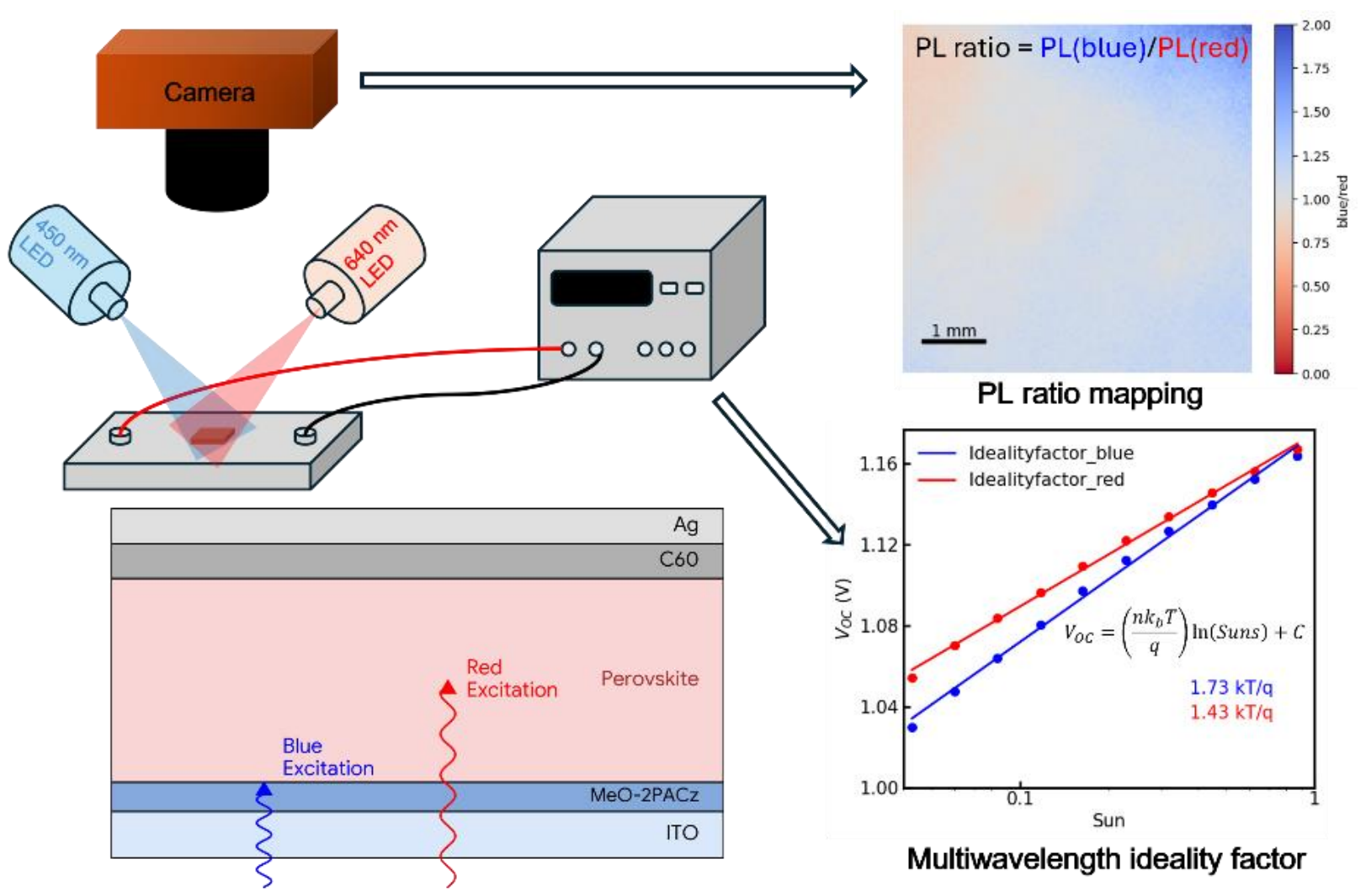

Metal halide perovskites are a promising candidate for next-generation photovoltaics, due to their high absorption coefficient, tunable optical bandgap and potential for additive manufacturing with low capital and materials costs.[1–5] To date, the power conversion efficiency of single-junction perovskite solar cells has increased to over 27%.[6] Despite the rapid progress in device efficiency, defects at metal halide perovskite interfaces can lead to trap-assisted nonradiative recombination, compromising device performance and long-term stability.[7–11] Notably, the use of phosphonic acids as interface modifiers at the buried interface has led to significant increases in record device performance for p-i-n perovskite cells.[12–16] As a result of this success, intense research interest has also focused on reducing non-radiative recombination at the perovskite/ETL interface, as a key remaining source of non-radiative voltage loss.[17–21]

While much effort has focused on improving device efficiency, long-term operational stability remains a critical challenge. Recent studies have shown that the buried HTL interface may remain a weak link for longevity, particularly under practical operating stresses such as heat,[22–24] reverse bias,[25] and UV illumination.[26–28] Although it is possible to monitor PCE over time under operational stress, such measurements provide limited information regarding the specific layers or interfaces responsible for device degradation. While a few studies have attempted to characterize degradation behaviors using depth-dependent measurements,[29–33] non-destructive and convenient methods for probing depth-dependent degradation under practical stressors remain an unmet need. This gap motivates the development of fast and non-destructive techniques for probing depth-dependent, interface-selective degradation in perovskite solar cells.

Photoluminescence (PL) imaging has been employed to map spatial variations in perovskite films and devices, providing fast and non-destructive measurements.[10,17,34,35] Correlated PL imaging enables a more detailed probing of perovskite film as well as solar cell device performance.[8,36–38] Prior work has employed multiwavelength photoluminescence (PL) as a non-destructive technique to probe depth-dependent information in tandem solar cells,[39–41] where the excitation wavelength is tuned to selectively probe different active layers based on their respective bandgaps. However, its application for single-junction perovskite solar cells is comparatively unreported,[42] and the potential for characterizing depth-dependent degradation under operational stresses offers an important and promising use case.

Recent work has reported strategies to improve the stability of perovskite solar cells under various operational stresses, including by strengthening the binding of organic HTLs to the TCO substrates or enhancing their intrinsic resistance to degradation.[26,43–49] However, systematically evaluating the effectiveness of such interface engineering strategies requires tools capable of isolating degradation at specific interfaces non-destructively — a capability that remains largely undeveloped.

UV illumination provides a particularly well-suited test case for demonstrating interface-selective, depth-resolved degradation characterization. For single-junction p-i-n PSCs, light is incident from the TCO/HTL side, and UV photons — owing to their short penetration depth — interact predominantly with this buried interface region. Under the AM 1.5G solar spectrum (100 mW/cm$^2$), UV light accounts for only ~4 mW/cm$^2$,[43] yet the high energy photons can generate deep trap states at interfaces, for example by oxidizing $I^-$ to $I_2$ and reducing $Pb^{2+}$ to Pb, and cause degradation or detachment of organic HTLs.[26,43,50–52] This tendency of UV-induced damage in p-i-n devices to localize to the HTL interface makes UV damage studies a useful model system for validating the ability of multiwavelength PL to resolve interface-specific degradation.

Herein we demonstrate a strategy employing multiwavelength excitation to characterize depth-dependent degradation in single-junction perovskite films and solar cells, using UV light soaking as a test case for interface-selective degradation. As an interfacial intervention, motivated by a recent report,[53] we mix a common silane-based interface modifier/adhesion promoter (3-mercaptopropyl)trimethoxysilane (MPTMS) (Figure S1) with the popular HTL modifier (2-(3,6-dimethoxy-9H-carbazol-9-yl)ethyl)phosphonic acid (MeO-2PACz) and test the silane's effect on UV stability. We use this modified interface as a model system to demonstrate the multiwavelength approach rather than as a new passivation strategy. We confirm that UV degradation primarily occurs at the HTL side using conventional characterization methods, and that MPTMS enhances the HTL stability on ITO substrate, leading to improved device UV resilience. By exploiting different excitation wavelengths with varying penetration depths, we perform depth-resolved multiwavelength wide-field PL (WFPL) and time-resolved photoluminescence (trPL) measurements on perovskite films and half stacks, and multiwavelength ideality factor measurements on solar cell devices. The results demonstrate that multiwavelength PL provides a fast and non-destructive tool for assessing depth-sensitive degradation — offering broad potential for systematic evaluation of interface engineering strategies across a range of operational stressors in future perovskite photovoltaics.

To investigate the effect of MPTMS on device passivation and UV stability, we first fabricate 1.67 eV bandgap p-i-n perovskite solar cells (Figure S2) with and without MPTMS incorporation at the TCO interface. This step was motivated by the report that mixing MPTMS with the phosphonic acid improves the performance and stability of the perovskite/ITO interface, which has been hypothesized to occur by preventing phosphonic acid HTL aggregation and passivating undercoordinated $Pb^{2+}$ at the perovskite/HTL interface.[43,50]

After fabrication, we characterized device performance both before and after exposure to UV light from the ITO side using a set of UV LEDs with a peak emission at 386 nm at a combined intensity of 33.2 mW/cm$^2$ (approximately ~7-8 times the UV irradiance in the AM 1.5G solar spectrum from 280-400 nm) for 20 hours. We carried out all subsequent UV stability experiments using the same illumination conditions. Figure 1a and 1b show the champion current density-voltage (*J*-*V*) curves of the devices with and without MPTMS treatment before and after UV exposure, respectively. The best performing MPTMS-treated device achieved a PCE of 21.24%, with a $V_{OC}$ of 1.16 V, $J_{SC}$ of 22.17 mA/cm$^2$, and FF of 81.88 %, showing an improvement over the control device (PCE = 19.68%, $V_{OC}$ = 1.14 V, $J_{SC}$ = 21.39 mA/cm$^2$, FF = 80.75%). After 20 hours of UV exposure, there is a significant decrease in PCE arising primarily from a substantial loss in $V_{OC}$ accompanied by reduced shunt resistance. By contrast, the MPTMS-treated device shows negligible change under the same conditions of UV exposure. Figure 1c – 1f show the statistics of a series of similarly prepared solar cell devices before and after UV treatment, confirming the enhanced UV stability of the MPTMS-treated devices. Table S1 summarizes the mean values and standard deviations of the device data. After UV soaking, the unpassivated devices show a significant PCE drop from 18.64% to 15.67%, driven primarily by $V_{OC}$ loss. By contrast, MPTMS inclusion effectively stabilizes the buried interface, with the treated devices retaining a PCE of 19.18% after the same UV exposure.

To study the impact of MPTMS on UV stability of phosphonic acid HTL modifiers and the perovskite, we also examine half-stacks of the structure ITO/HTL/perovskite. First, we employ attenuated total reflection infrared (ATR-IR) spectroscopy to these two HTLs on ITO glass to probe the evolution of interfacial bonding upon UV treatment. Figure S3a and S3b show the ATR-IR absorption spectra of the treated interfaces before and after UV soaking, respectively. In both spectra, we attribute the signal peaks at 1213 and 1162 $cm^{-1}$ to the vibrational modes of P=O and P-O-H of MeO-2PACz.[43,54] For the MeO-2PACz control (Figure S3a), both signal peaks decrease after 20 hours of UV treatment, suggesting desorption of the HTL and subsequent degradation of the buried interface, consistent with previous reports of UV exposure to phosphonic modified ITO.[43]

In contrast, with the addition of MPTMS, we observe the ATR-IR absorption spectra are virtually unchanged (Figure S3b) following the same period of UV exposure. While the exact path of MPTMS action is not the focus of this work, these data provide new mechanistic insight, suggesting that mixing MPTMS into the phosphonic acid HTL improves stability by somehow preventing UV-induced detachment of the phosphonic acid from the ITO.

To further study the effect of MPTMS treatment of the buried interface on the UV stability of perovskite half-stacks, we also perform XRD measurements to monitor the perovskite phase stability under UV exposure. In figure S4, the half-stack deposited on a mixed MPTMS/MeO-2PACz treated surface initially shows a lower signal of $PbI_2$ (peak at ~ 12.7°)[53] compared to the MeO-2PACz reference half-stack. After 20 hours of UV treatment, the XRD peak corresponding to $PbI_2$ significantly increases for the control half stack without MPTMS. By contrast, the MeO-2PACz + MPTMS half stack shows minimal change in the $PbI_2$ peak intensity following UV exposure, suggesting that either the MPTMS itself, or the reduced desorption of the phosphonic acid, helps prevent perovskite phase degradation under UV exposure.

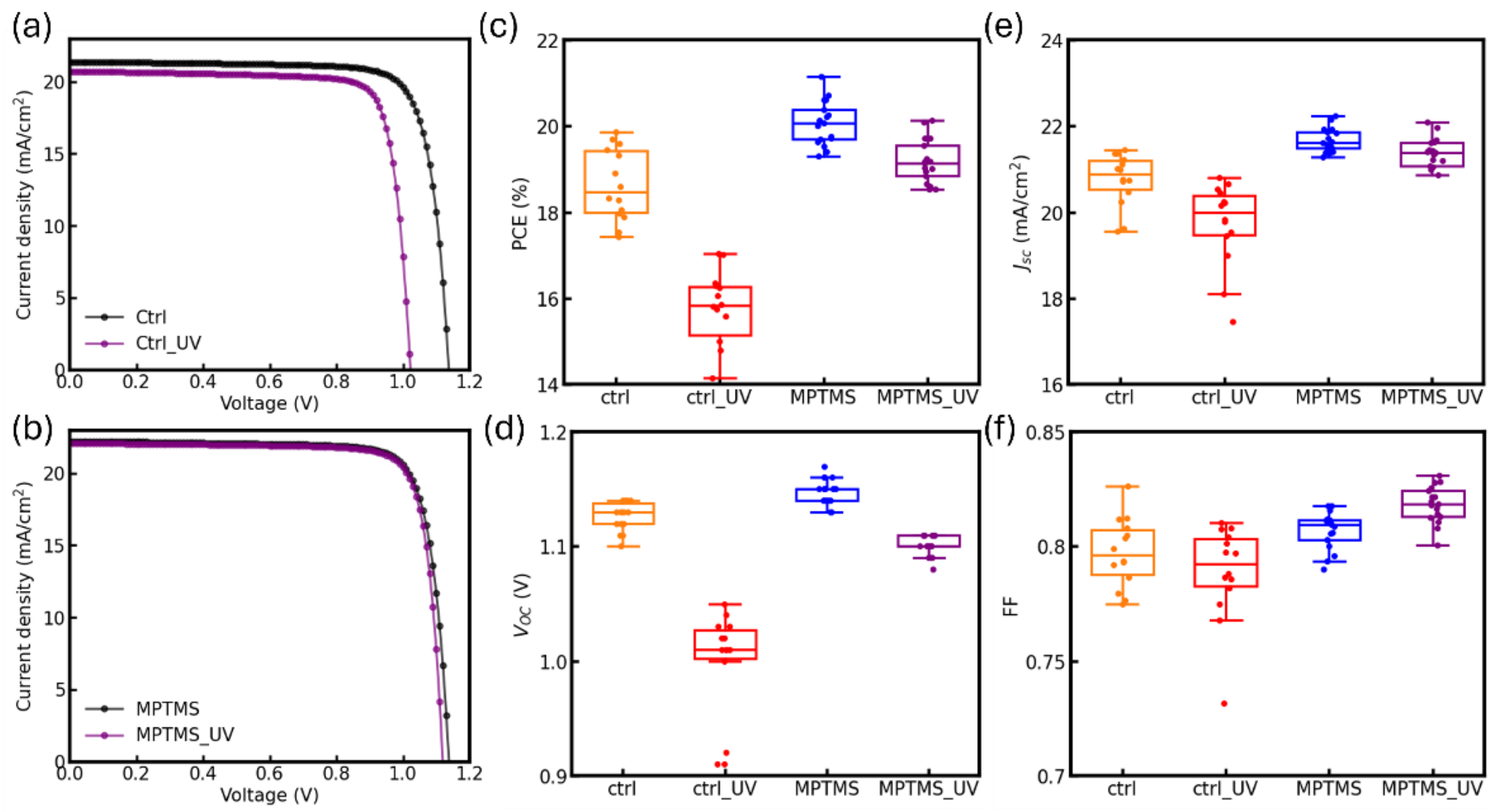

**Figure 1.** Champion *J-V* curves of cells with (a) MeO-2PACz-modified HTL interface and (b) MeO-2PACz + MPTMS modified HTL interface solar cell. Boxplots from N = 6 total devices for (c) power conversion efficiency (PCE), (d) open-circuit voltage ($V_{OC}$), (e) short-circuit current density ($J_{SC}$), and (f) fill factor of MeO-2PACz only and MeO-2PACz + MPTMS solar cells both before and after 386 nm UV exposure at 33.2 mW/cm$^2$ for 20 hours.

Having characterized the UV-induced degradation and stability of the perovskite buried interface using conventional techniques, we next explore a fast, convenient, and non-destructive multiwavelength photoluminescence (PL) strategy to probe depth-dependent degradation in single-junction perovskite solar cells. Multiwavelength PL has previously been applied in tandem solar cells to access different active layers by the use their different bandgaps.[39,40,55] Here, we extend this concept to single-junction devices by exploiting the wavelength-dependent optical penetration depth.[42]

Perovskite materials exhibit strong wavelength-dependent optical absorption.[42] Shorter-wavelength light is absorbed more strongly, and this generates more carriers near the film surface, whereas longer-wavelength light penetrates deeper into the perovskite films. Figure S5 shows the absorption coefficient spectrum and calculated optical penetration depth as a function of wavelength for the perovskite film with the same composition we use here, measured by spectroscopic ellipsometry. We calculated the optical penetration depth from the absorption coefficient using the standard relation $\alpha = 4\pi k/\lambda$, where the penetration depth is defined as $1/\alpha$, corresponding to the depth at which the light intensity decays to 1/e of its value at the surface.

The depth-dependent photocarrier generation rate is given by the spatial derivative of the absorbed photon flux:

$$G(z) = -\frac{dI(z)}{dz} = \alpha I_0 e^{-\alpha z}$$

Figure 2a shows the resulting depth-dependent photocarrier generation profiles, $G(z)$, under 450 nm and 660 nm excitation. Under 450 nm excitation, photogeneration peaks at the surface and decays rapidly with depth ($\alpha_{450\,nm} = 2.12 \times 10^5$ cm$^{-1}$), while 660 nm ($\alpha_{660\,nm} = 4.29 \times 10^4$ cm$^{-1}$) excitation produces a relatively more uniform carrier distribution throughout the perovskite layer. While the photogenerated carriers can diffuse due to concentration gradients and drift under applied fields, this level of contrast can be sufficient to distinguish surface and bulk phenomena.[42]

Exploiting this wavelength-dependent absorption, we tune the excitation wavelength to modulate the depth profile of photogenerated carriers, enabling depth-sensitive photoluminescence measurements of UV-induced degradation in single-junction perovskite films. We next demonstrate multiwavelength wide-field PL imaging capability for both spatially and vertically-resolved characterization of perovskite films as well as solar cell devices (Figure 2b). For the multiwavelength wide-field PL imaging setup, we use a scientific CMOS camera (Thorlabs CS895MU) to record low-intensity near-infrared photoluminescence images from perovskite films and devices with a centimeter-scale imaging area. In this experiment, we select collimated LEDs with peak wavelengths of 450 nm and 660 nm as excitation sources, which preferentially excite more surface or less-surface dominated regions.

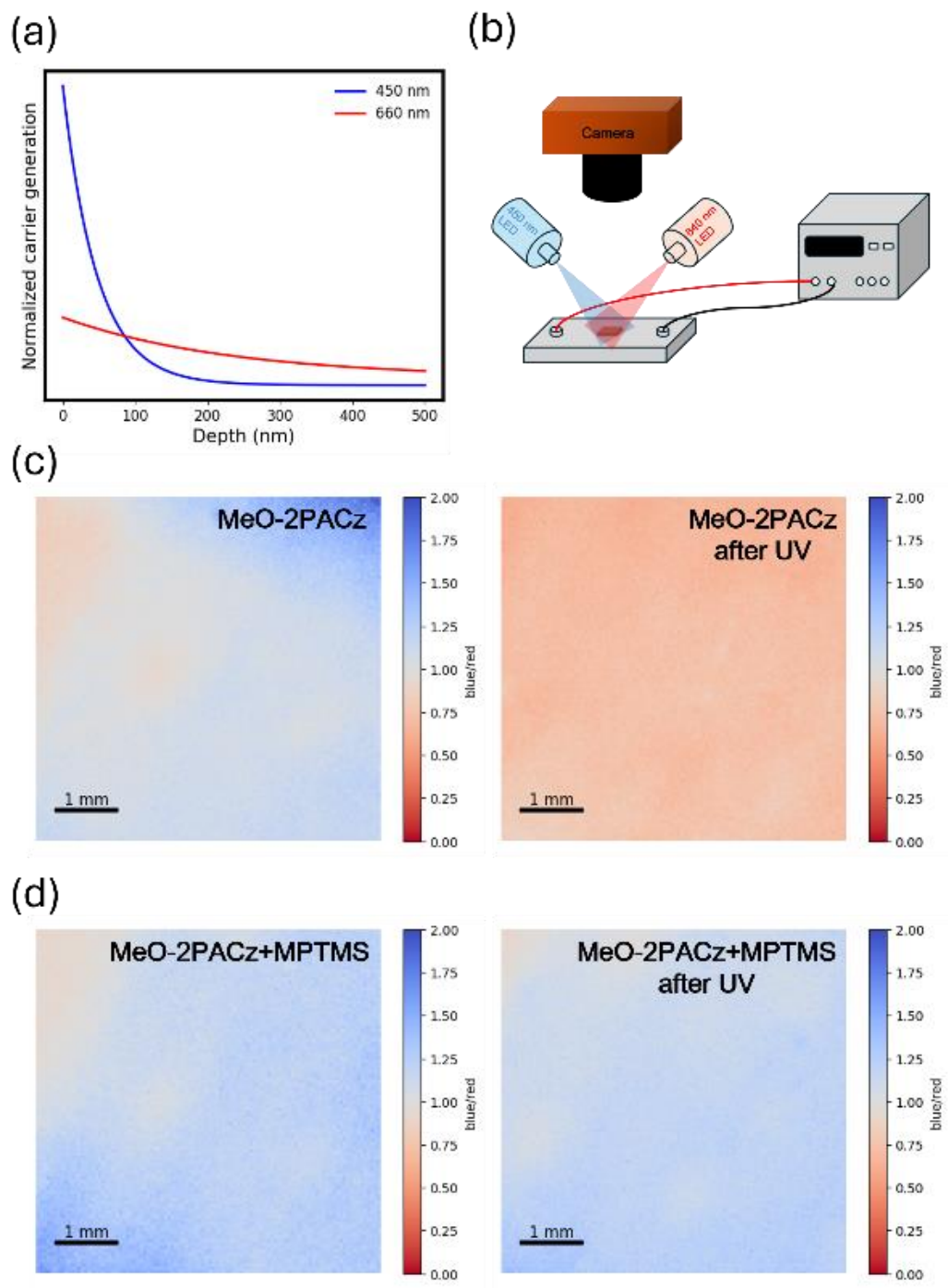

**Figure 2.** (a) Calculated depth-dependent photocarrier generation profile G(z) in the perovskite film under 450 nm and 660 nm excitation. (b) Schematic of multiwavelength wide-field image setup. Images are taken using the setup described in the SI in full detail. Ratiometric images showing the intensity of perovskite IR PL resulting from blue (450 nm) excitation, divided by the perovskite IR PL resulting from red (660 nm) excitation on the same area for (c) MeO-2PACz-treated ITO half-stacks before and after UV (386 nm 33.2 mw/cm$^2$ for 20 hours) exposure; along with (d) MeO-2PACz + MPTMS-treated ITO half stacks before and after UV exposure.

Importantly, with this PL imaging experiment, in addition to probing any lateral heterogeneity as is typical for PL mapping, we can also distinguish PL coming from more surface dominated and deeper/bulk regions by using different wavelengths excitation. Figure S7, shows the PL images with blue (450 nm) and red (660 nm) excitation on the same area, with fluences set to match the corresponding device photocurrent density under 1 sun (~21.0 mA/cm$^2$). By simply dividing the blue-excited PL by red-excited PL, we can obtain a ratiometric image (Figure S8) to qualitatively compare the carrier recombination between the surface and bulk regions of the perovskite, which reveals depth-dependent heterogeneity across the perovskite film.

Figure 2c and 2d show ratiometric PL images (blue-excited/red-excited) for MeO-2PACz and MeO-2PACz + MPTMS half-stacks before and after UV soaking, respectively. In Figure 2c, ratio mapping before UV treatment shows an initially higher PL under blue excitation. After 20 hours of UV soaking, ratio mapping significantly changes from "blue" to "red", together with the decrease of PL counts in Figure S9 and S10, indicating the HTL side undergoes more UV degradation than the perovskite bulk for the control half-stack. In contrast, Figure 2d shows a blue-excited/red-excited PL ratio mapping after UV soaking that is similar to the unaged ratio mapping of the MPTMS-treated half-stack, suggesting a more UV-stable buried interface with MPTMS treatment, which is consistent with the previous conventional measurements. We plot the histograms in figure S13 and S14 to compare the bulk and surface PL change after UV soaking. The control half-stack (figure S13) shows a more significant decrease in PL when excited by 450 nm LED from the HTL side, suggesting a more degraded buried interface following UV stress. In contrast, for MPTMS-treated half-stack, the dominant interfacial degradation is effectively suppressed, leading to a more uniform but mild UV degradation.

To corroborate the above observations from multiwavelength PL mapping, we further explore multiwavelength trPL to probe the depth-dependent carrier recombination dynamics under UV exposure. Figure 3a shows the confocal multiwavelength trPL set up, where two pulsed laser sources with different excitation wavelengths (470 and 640 nm) are focused onto the perovskite/HTL/ITO half-stack through a microscope objective. The emitted photoluminescence is collected by the same objective, spectrally separated by a dichroic beam splitter, and directed to a time-resolved detector for carrier lifetime analysis.

Figure 3b and 3c show the PL decays of the control and MPTMS-treated half-stacks, respectively, excited from the HTL side before and after UV soaking, with laser fluences adjusted to equal photon absorption per pulse. Table S2 summarizes the fitted lifetime and parameters of all decays. Before UV soaking, the MPTMS-treated half-stack exhibits longer carrier lifetimes under both excitation wavelengths, with the average 470 nm lifetime increasing from 87 ns (control) to 158 ns (MPTMS), again consistent with including MPTMS treatment reducing recombination losses at the HTL side. After 20 hours of UV stress, the control half-stack shows severe degradation at the HTL side, with the average blue-excited lifetime dropping from 87 ns to 36 ns, while the MPTMS-treated sample shows a comparatively mild and uniform lifetime decrease under both excitation wavelengths. Taken together, the multiwavelength PL imaging and confocal trPL measurements consistently confirm that UV-induced degradation preferentially occurs at the HTL interface, and that MPTMS treatment effectively mitigates this interfacial damage, in agreement with the prior results from conventional characterizations.

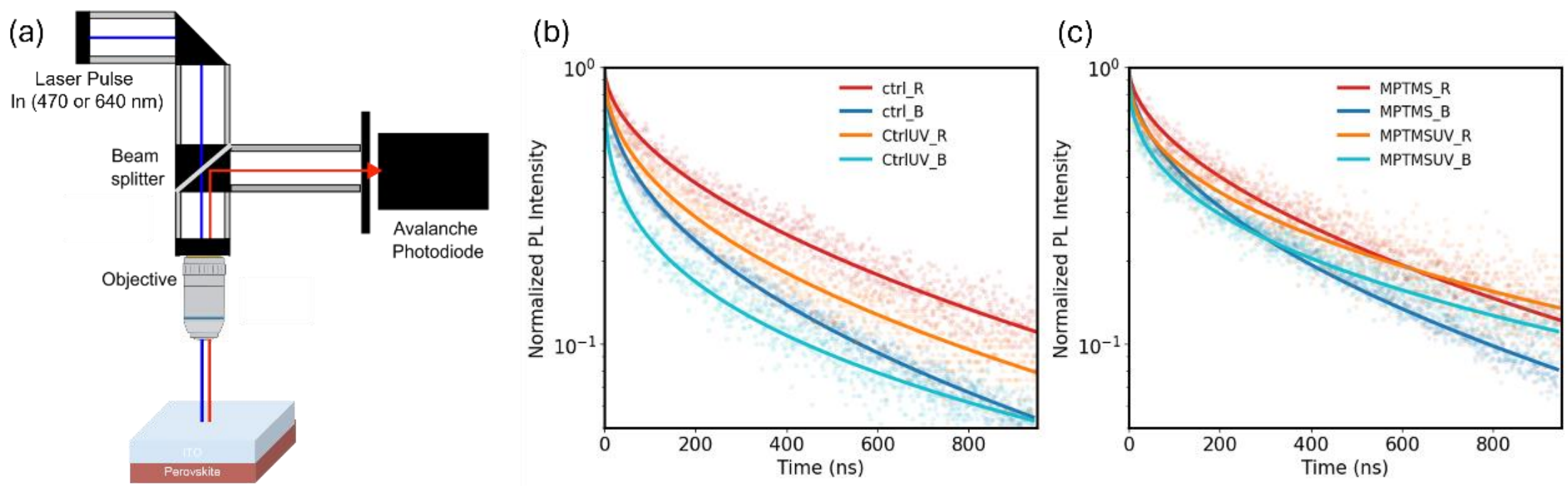

**Figure 3.** (a) Schematic of confocal multiwavelength time-resolved PL set up. PL decays of (b) untreated and (c) MPTMS-treated perovskite half-stack excited at the back side with red (640 nm) and blue (470 nm) lasers before and after UV treatment. Untreated half-stack has a more pronounced lifetime reduction, particularly under blue laser excitation (88 ns to 33 ns), while the MPTMS-treated sample shows a higher initial lifetime and a comparatively milder lifetime reduction after UV treatment.

Having established depth-dependent UV degradation in perovskite half-stacks using multiwavelength PL imaging and trPL, we extend this approach to full solar cell devices using a generalization of device ideality factor analysis we term multiwavelength ideality factor measurements. The ideality factor $n_{id}$ describes how much a device deviates from ideal diode behavior. It is obtained from the dependence of open-circuit voltage on light intensity:

$$V_{OC} = \frac{n_{id}kT}{q}\ln(I) + C$$

Conventionally, ideality factors are often interpreted in terms of idealized recombination mechanisms. In textbook treatments, $n_{\mathrm{id}} = 1$is commonly associated with second-order bimolecular recombination of free carriers, including radiative band-to-band recombination, whereas $n_{\mathrm{id}} = 2$ is commonly associated with first-order monomolecular recombination, such as trap-assisted Shockley–Read–Hall recombination through deep trap states.

In real solar cells, however, the measured ideality factor can exhibit more complicated behavior, especially when interfacial recombination, transport limitations, trap filling, or ionic effects influence the measured $V_{OC}$.[56]

Nevertheless, the ideality factor measurement is widely employed as a diagnostic tool for solar cell devices, and is relatively straightforward to perform at a scale from cells to modules.

Next, by performing ideality factor measurements under 450 nm and 660 nm excitation using the same wide-field setup, with the device connected to a source meter to record $V_{OC}$ in place of the camera — we show that the ideality factor can encode depth-sensitive information, enabling us to distinguish recombination losses from the perovskite/HTL interface versus the perovskite bulk.

Figure 4a and 4b show the multiwavelength ideality factors of the control MeO-2PACz device before and after UV soaking. UV soaking leads to a decrease in $V_{OC}$ and an increase in ideality factor. Importantly, both changes become more pronounced under blue (450 nm) excitation than red (660 nm) excitation, indicating that UV-induced defect-assisted recombination is concentrated at the buried HTL interface rather than the bulk. In contrast, figure 4c and 4d show that the MPTMS-treated device exhibits significantly smaller changes in both $V_{OC}$ and ideality after UV treatment under both excitation wavelengths, confirming that MPTMS passivation effectively suppresses interfacial degradation at the device level. These device-level results are consistent with the half-stack PL and trPL measurements, indicating that UV degradation mainly occurs at the HTL side of the solar cell, and the optimization of the buried interface can enhance the device's UV stability.

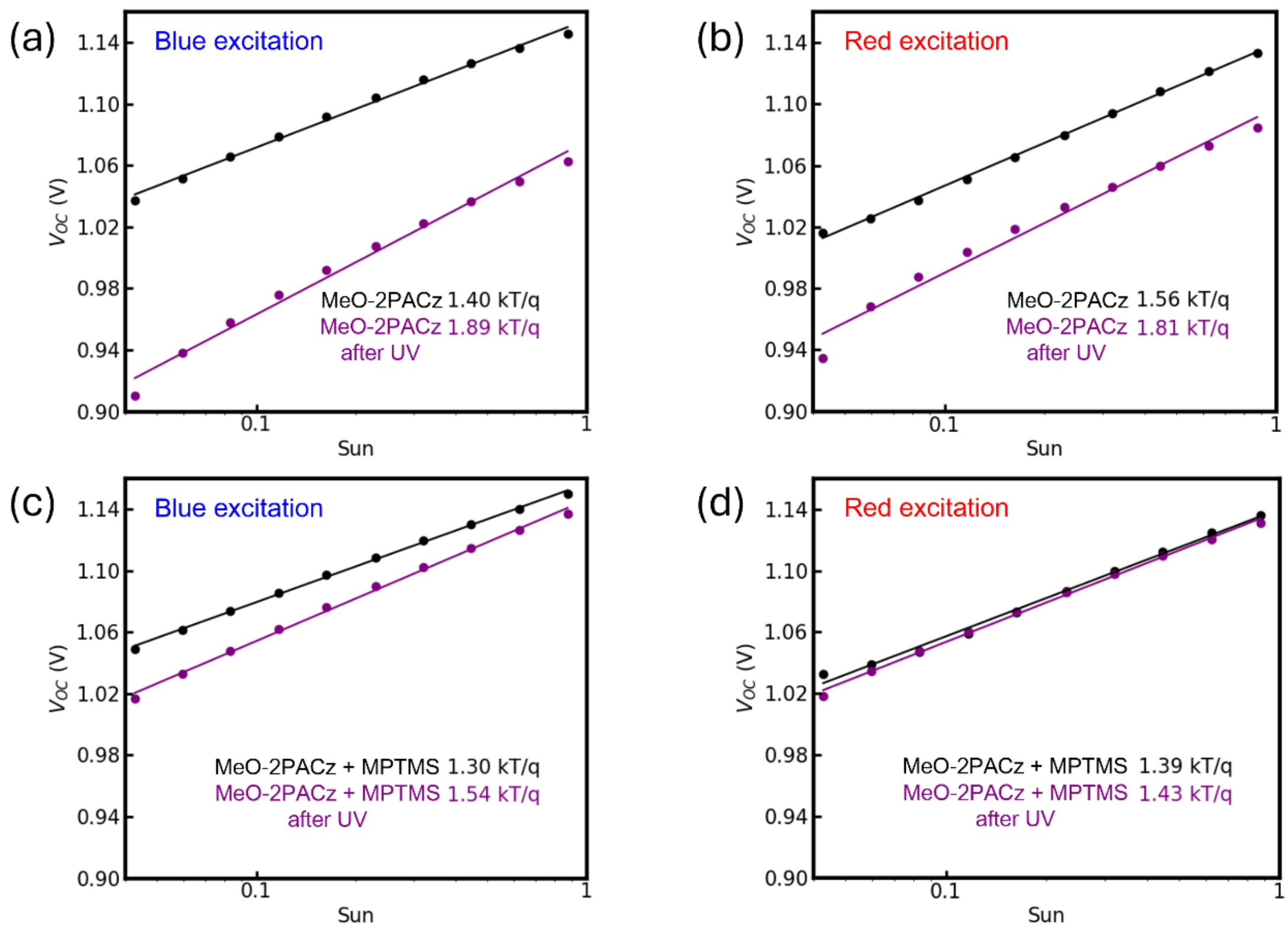

**Figure 4.** Ideality factors of control perovskite device with (a) blue and (b) red excitation. After 20 hours of UV treatment, the ideality factor increases and $V_{OC}$ decreases noticeably, particularly under blue excitation, with $V_{OC}$ at maximum operating intensity dropping from 1.145 V to 1.062 V, and ideality factor increases from 1.40 to 1.89 after UV treatment. Ideality factors of MPTMS-treated devices with (c) blue and (d) red excitation before and after UV treatment, showing comparatively smaller decrease in $V_{OC}$ (from 1.147 V to 1.139 V with blue excitation at maximum operating intensity) and smaller increase in ideality factor (from 1.30 to 1.54 with blue excitation).

In summary, we demonstrate that wide-field steady-state PL, trPL and ideality factor measurements can probe the depth-dependent degradation behaviors under UV stressing of both perovskite films and single-junction solar cell devices using wavelength-dependent excitation. With this approach, we confirm that UV-induced degradation occurs preferentially from the ITO (HTL) side, and a buried-interface treatment with MPTMS enables a better UV stability of the perovskite solar cell. These results are self-consistent with conventional characterization including ATR-IR and XRD, and with previous reports. Importantly, this study provides a fast and non-destructive approach to assess the depth-dependent recombination and degradation occurring under operando stress, here using UV soaking as a test case. We anticipate this approach can enable faster identification of vulnerable layers and evaluation of degradation pathways of perovskite solar cells under practical stressors while providing insights for improving device stability in future perovskite industries.

**Acknowledgment**

This material is based upon work supported by the U.S. Department of Energy's Office of Energy Efficiency and Renewable Energy (EERE) under the Solar Energy Technologies Office Award Number DE-EE0011422. The views expressed herein do not necessarily represent the views of the U.S. Government or any agency thereof. The authors acknowledge the use of facilities and instruments at the Photonics Research Center (PRC) at the Department of Chemistry, University of Washington; Research Training Testbed (RTT), part of the Washington Clean Energy Testbeds system. Part of this work is conducted at the Molecular Analysis Facility (MAF), a National Nanotechnology Coordinated Infrastructure site at the University of Washington which is supported in part by funds from the National Science Foundation (awards NNCI-2025489, NNCI-1542101), the Molecular Engineering & Sciences Institute. D.S.G. and Z.H. acknowledge support from the Washington Research Foundation and Rabinovitch Endowment for completing the manuscript following the termination of the award.